\documentstyle[11pt,epsf]{article}
\input epsf
\begin{document}
\baselineskip 100pt
\renewcommand{\baselinestretch}{1.5}
\renewcommand{\arraystretch}{0.666666666}
{\large
\parskip.2in
\newcommand{\be}{\begin{equation}}
\newcommand{\ee}{\end{equation}}
\newcommand{\br}{\bar}
\newcommand{\fr}{\frac}   
\newcommand{\lm}{\lambda}
\newcommand{\ra}{\rightarrow}
\newcommand{\al}{\alpha}
\newcommand{\bt}{\beta}
\newcommand{\pr}{\partial}
\newcommand{\hs}{\hspace{5mm}}
\newcommand{\up}{\upsilon}
\newcommand{\dg}{\dagger}
\newcommand{\ve}{\varepsilon}
\newcommand{\acc}{\\[3mm]}
\newcommand{\dl}{\delta}

\hfill DTP\,98/09

\bigskip
\begin{center}
{\bf Lagrangian Formulation of the General Modified Chiral Model}
\end{center}
\bigskip

\begin{center}
T{\small HEODORA} I{\small OANNIDOU} and W{\small OJTEK} Z{\small
AKRZEWSKI}\\
{\sl Department of Mathematical Sciences, University of Durham,\\
Durham DH1 3LE, UK}
\end{center}

{\bf Abstract.}
We  present a Lagrangian formulation for the general modified
chiral model.  
We use it to discuss the Hamiltonian formalism for this model and to
derive the commutation relations for the chiral field. We look at some
explicit examples and show that the Hamiltonian, containing a contribution
involving a Wess-Zumino term, is conserved, as required.

\vskip 0.5cm
In recent years, the modified chiral model in (2+1)  dimensions has
emerged as a beautiful and powerful mathematical system, possessing
soliton solutions with trivial \cite{W} and non-trivial scattering
properties\cite{W,ola,IZ2}.
The model possesses many traits of integrable systems such as a linear
system \cite{W} and an infinite number of conservations laws \cite{IW}.
All its studies so far have been purely classical. The time has come 
to start developing a quantum version of the model. To do this we have, 
first of all, to find the action of the model and then proceed 
to its quantisation.

To find the action  we note that usually one thinks of the 
 action as given by a space-time integral of a local
Lagrangian density, whose variation gives the classical equations of
motion.
 However, since the modified chiral  model has been derived
by a dimensional reduction of the
self-dual
Yang-Mills equations in (2+2) dimensions, there is no known Lagrangian 
formulation for it.
Therefore, to have an action we have to include
a contribution of the WZW type \cite{Wit2}.
A similar situation arises when we want to have an effective theory of OCD
\cite{Wit1} or derive a bosonized description of free fermions in two
dimensions \cite{Wit2}.

In this letter we present a similar description of the
general modified chiral model \cite{IZ}, {\it ie} whose equation 
of motion is
\be
(\eta^{\mu \nu}+\ve^{\mu \nu \al}V_\al) \pr_\mu(\pr_\nu \Psi
\Psi^\dg)=0.
\label{ch}
\ee
Here, $\Psi$ is a map from ${\bf R}^{2+1}$ to $SU(2)$ which can be
thought of as a $2 \times 2$ unitary matrix valued function of coordinates
$x^\mu=(t,x,y)$ and where $^\dagger$ denotes the hermitian conjugation.   
Greek indices range over the values $0, 1, 2$, $\pr_\mu$ denotes partial
differentiation with respect to $x^\mu$,
$\ve^{\mu \nu \al}$ is the totally skew tensor with $\ve^{012}=1$, and
$V_\al$ is a constant unit vector.
Indices are raised and lowered using the Minkowski metric $\eta^{\mu 
\nu}=\mbox{diag}(-1,1,1)$ and $V_\al$ is of the form $V_\al=(\lm, 1, \lm)$.
Thus  $\Psi$, a solution of (\ref{ch}), is a function of $\lm$. Note that
if we set
$\lm=0$ (\ref{ch}) reduces to  Ward's model \cite{W} and $\Psi$
is independent of $\lm$.   

As we have said above  (\ref{ch}) cannot be derived from a standard Lagrangian.
To see this we note that we would need something like $V_\al \ve^{\mu
\nu \al} \mbox{tr}(\pr_\mu\Psi\Psi^\dg\, \pr_\nu\Psi\Psi^\dg)$ but this
vanishes, by antisymmetry of $\ve^{\mu \nu \al}$ and the cyclic symmetry of the
trace.
Nevertheless, we can find an action for  (\ref{ch}) which 
involves the usual Lagrangian term and a further WZW-like term. 
This modified action will be then compared with the Hamiltonian version of
the model.
To derive them we  follow the procedure developed in \cite{CY}
 for the self-dual Yang
Mills system; thus in addition to the usual term we consider 
also a  Wess-Zumino action (term) \cite{WZ}, {\it ie}
\be
S\!=\!-\fr{1}{2}\!\int_{R^2}\!\!dx\,dy\!\int_{t_1}^{t_2}\!\!dt\,
\mbox{tr}(\pr_\mu\Psi^\dg  \pr^\mu \Psi)\!+\!\fr{1}{3}
\!\int_{R^2}\!\!dx\,dy\!\int_{t_1}^{t_2}\!\!dt\!\int_0^1\!\!d \rho
\,V_i\,\ve^{ijkl} \,\mbox{tr}(\tilde{\Psi}
\pr_j\widetilde{\Psi}^\dg \, \widetilde{\Psi} \pr_k\widetilde{\Psi}^\dg
\,\widetilde{\Psi} \pr_l\widetilde{\Psi}^\dg).
\label{act}
\ee
In the second  term in ({\ref{act}}) Latin indices range over the values
$0,1,2,3$, with $x^3=\rho$ and $V_i=(\lm,1,\lm,0)$.
Thus our action (\ref{act}) is a sum of the action of the
principal chiral model and of the Wess-Zumino term, which is locally (but not
globally) a total divergence.

The actual form of our
three dimensional Wess-Zumino term can be derived by analogy with
a similar treatment in four dimensions.
To do this the unitary matrix $\Psi$ has to be extended to
$\widetilde{\Psi}$, a function of $(t,x,y,\rho)$, where the additional
variable $\rho$ satisfies $0\le \rho\le 1$.
So, working in Minkowski space, we imagine our space-time to be the surface of
a cylinder ${\bf S^2}\times \bf{R}$ with an infinitely large base defined 
by $(x,y)$ and height $t$.
Our extended mapping can be chosen in any way which satisfies
the appropriate boundary conditions; a convenient way is to consider it
as an extension of the mapping
 $\Psi$ from  ${\bf R}^{2+1}$ to $SU(2)$  to the
mapping $\widetilde{\Psi}$ of the interior of the cylinder, {\it ie}
\be
\widetilde{\Psi}(t,x,y,\rho)=\left[ \begin{array}{ll}
\Psi(t,x,y), & \hs \rho=1,\acc
\Psi_0=\mbox{const}, & \hs \rho=0.
\end{array} \right.
\ee
To check that we have a correct action we verify that  (\ref{act}) reproduces
our equations of motion (\ref{ch}).
First we note that the variation of the Wess-Zumino term is a simple local
functional.
Then we find that the variation of the action $S$ is given by
\be
\!\!\!\delta 
S\!\!=\!\!\int_{R^2}\!\!dx\,dy\!\int_{t_1}^{t_2}\!\!dt\,\mbox{tr}\left(\Psi
\delta \Psi^\dg \,\eta^{\mu\nu}\pr_\mu(\pr_\nu\Psi\Psi^\dg)\right)\!\!+
\!\!\int_{R^2}\!\!dx\,dy\!\int_{t_1}^{t_2}\!\!dt\,\mbox{tr}\left(\Psi
\delta
\Psi^\dg \, \ve^{\mu\nu\al}\,V_\al\pr_\mu(\pr_\nu\Psi \Psi^\dg)\right).
\ee
The variational equation  is therefore (\ref{ch}), assuming that the
variation of the fields at the boundaries vanishes.

Next we derive the canonical formalism for the model. To do this we note
that we need a version of the theory
with an action  that is first order in time derivatives. 
We note that the new (WZW) part of (\ref{act})  is {\it already} in 
the Hamiltonian form; that is, it is
already of the first order in time derivatives. Moreover, as 
it contains first-order time
derivatives, it does not contribute to the Hamiltonian. 

To proceed further it is convenient to imagine the $\rho$ integration,
for the terms linear in time derivatives in the WZW part of the action, as
 having been
carried out, and so to rewrite the action as
\be
S\!=\!\fr{1}{2}\!\int_{R^2}\!\!dx\,dy\!\int_{t_1}^{t_2}\!\!dt\left\{\mbox{tr}
\left(\pr_\mu \Psi^\dg \pr^\mu \Psi\!+\!2A[\Psi^\dg]\pr_t\Psi^\dg\right)
\!+\!2\lm\!\!\int_0^1\!\!d\rho\,\mbox{tr}\left(\pr_\rho\widetilde{\Psi}^\dg
\widetilde{\Psi} [\pr_x\widetilde{\Psi}^\dg\widetilde{\Psi},
\pr_y\widetilde{\Psi}^\dg\widetilde{\Psi}]\right)\right\},
\label{tel}
\ee
with $A[\Psi^\dg]$ some unknown matrix-valued function of $\Psi^\dg$.
Note that the action also contains terms independent of the time
derivatives; such
terms do not contribute to the Poisson brackets.

As in  \cite{Wit2,Abd} we note that although
the form of $A$ is rather complicated, fortunately, we do not need to
know  $A[\Psi^\dg]$ itself, but only the
antisymmetric tensor
\be
F_{ij,kl}=\fr{\pr A_{ij}}{\pr \Psi^\dg_{lk}}-\fr{\pr A_{kl}}{\pr
\Psi^\dg_{ji}}.
\ee
This tensor is explicitly calculable by comparing the variations of the
$t$-dependent Wess-Zumino term with respect to
$\Psi^\dg$ as calculated from (\ref{act}) and (\ref{tel}), respectively.
After some calculation, we find
\be
F_{ij,kl}=\Psi^{kj}(\pr_y\Psi-\lm\pr_x\Psi)^{il}
-(\pr_y\Psi-\lm\pr_x\Psi)^{kj}\Psi^{il}.
\ee

Considering the Lagrangian density of (\ref{act}), the momentum conjugate
to $\Psi_{ij}$ is  given by
\be
\Pi_{ij}=\pr_t\Psi_{ij}+A_{ij}.
\label{P}
\ee
Since, as we have already mentioned, the Hamiltonian does not depend on
the unknown function $A[\Psi^\dg]$, it is convenient to define new
momentum variables by
\be
\widetilde\Pi_{ij}\equiv \pr_t\Psi_{ij}=\Pi_{ij}-A_{ij}.
\ee
It then follows that the Hamiltonian is given by 
\be
H\!\!=\!\!-\fr{1}{2}\int_{R^2}\!\!dx\,dy\left\{ 
\mbox{tr}\left((\Psi^\dg\widetilde\Pi)^2
\!+\!(\Psi^\dg\pr_x\Psi)^2\!+\!(\Psi^\dg\pr_y\Psi)^2\right)
\!-\!2\lm\!\!\int_0^1\!\!d\rho \,\mbox{tr}\left(
\pr_\rho\widetilde{\Psi} \widetilde{\Psi}^\dg[
\pr_x\widetilde{\Psi} \widetilde{\Psi}^\dg,
\pr_y\widetilde{\Psi} \widetilde{\Psi}^\dg]\right)\right\}.
\label{ham}
\ee
The Hamiltonian equations of motion, written in terms of Poisson brackets,
are
\begin{eqnarray}
\label{p}
\pr_t\Psi^\dg&=&\{H,\Psi^\dg\},\\
\label{up}
\pr_t\widetilde{\Pi}&=&\{H,\widetilde{\Pi}\},
\end{eqnarray}
where the Poisson brackets on  $\Omega$ ({\it ie} the space of matrix
functions of $(x,y)$) are defined \cite{W1}
\be
\{f,g\}=\fr{\dl f}{\dl \widetilde{\Pi}}\bullet\fr{\dl g}{\dl
\Psi^\dg}-\fr{\dl
f}{\dl \Psi^\dg} \bullet\fr{\dl g}{\dl \widetilde{\Pi}}
\ee
and the scalar product on $\Omega$ is defined as
$A\bullet B=\int_{R^2} dx\,dy\,\mbox{tr}(AB)$.

Using the properties
\be
\{\Psi^\dg_{ij},\widetilde\Pi_{kl}\}=\dl_{ik}\dl_{jl},\hs \hs
\{\widetilde\Pi_{ij},\widetilde\Pi_{kl}\}=-F_{kl,ij},
\ee
we observe that the equation (\ref{p})  reduces to (\ref{P}), 
while the equation (\ref{up}),
after substitution of (\ref{P}), leads to the equation of motion
(\ref{ch}).
We readily verify that the explicit knowledge of the function $A[\Psi^\dg]$
is not needed.

As has been shown  in \cite{IZ2}, solutions of (\ref{ch}) correspond to 
products of factors called $t$-dependent unitons.
Thus it is interesting to compute the values of the energy for
some of these solutions; and, especially, look at the 
contribution of the extra term in
the Hamiltonian.
To do so, we have to find a convenient way of computing explicitly the
contribution  of the Wess-Zumino term.
To perform this explicitly we, first of all, consider a 1-uniton field
configuration $\Psi=(1-aR)$, where $a=2/(1+\lm)$.
For our extension we can now use
\begin{eqnarray}
\widetilde{\Psi}(t,x,y,\rho)&=&e^{ib\rho R}\\
&=&1+(e^{ib \rho}-1)R,
\end{eqnarray}
where
\be
b=\arctan{\fr{2\lm}{\lm^2-1}}.
\label{b}
\ee
For this solution, using the complex coordinates and the equations
satisfied by the projector $R$, {\it ie}
\be
 \pr_+R R=0,
\label{eRe}
\ee
 we get
\be
E_{1-\mbox{\small un}}=(8-2\lambda b)
\int_{R^2}d^2x\,\mbox{tr}(\pr_+R\pr_-R).
\label{ene}
\ee
It is important to note that $b$ is defined only up to the addition of
$2\pi$ and so, the value of the energy (\ref{ene}) is not uniquely
defined; {\it ie} it depends on the parameter $\lm$.
This comes from the possibility of choosing a different extension of $\Psi$,
{\it ie} which corresponds to a different value of the arbitrary constant $b$.

The computation of the value of the energy for more general solutions is
 more complicated. This is because now we have to rely 
on the continuation involved in the definition of the 
Wess-Zumino contribution.
Thus, if we consider, for example, a solution corresponding to a 2-uniton
field, we have $\Psi=(1-aP)(1-aR)$ where $a=2/(1+\lm)$.
Note that for $\lm=0$, the energy of this solution is the energy of the
principal chiral model in (2+1) dimensions which, as is well 
known, is conserved and can be
computed with ease (cf. \cite{W}-\cite{IZ2}).
For the nonvanishing values of $\lm$, our extension becomes
\begin{eqnarray}
\widetilde{\Psi}(t,x,y,\rho)&=&Ke^{ib\rho P}e^{ib\rho R}\nonumber\\
&=&K\left(1+(e^{ib \rho}-1)P\right)\left(1+(e^{ib \rho}-1)R\right),
\end{eqnarray}
where $b$ is given by (\ref{b}), $K$ is a constant $SU(2)$ matrix.
For this solution we can use the equations satisfied by the projector $P$ 
\be   
\pr_t P\, P+2 i(1-P)\,\pr_-R \,P=0,\hs \hs \hs P\,\pr_- P+P\,\pr_-R
\,(1-P)=0.
\label{gene}
\ee
to find that
\begin{eqnarray}
E_{2-\mbox{\small un}}=&&\!\!\!\!\!\!\!\!\!\!(8-4\lm b)
\!\!\int_{R^2}\!d^2x\,\mbox{tr}\left(\pr_+R\pr_-R\!
+\![\pr_+P,\pr_-P](R\!+\!P)\!+\![\pr_+R,\pr_-R]P
\!-\!P\pr_+P \pr_-R\!-\!P\pr_+R \pr_-P\right)\nonumber\acc
&&\!\!\!\!\!\!\!\!\!\!+\fr{8}{\lm^2+1}\int_{R^2}\!d^2x\,
\mbox{tr}\left(P\pr_+R\pr_-R+
P\pr_-R\pr_+R-2P\pr_+RP\pr_-R-[\pr_+P,\pr_-P]R\right).
\end{eqnarray}

It is easy to check that $E_{2-\mbox{\small un}}$ is, 
as required, a constant of motion.
Its $\lm$ dependence resides entirely in the prefactors
$(8-4\lm b)$ and $8/ (\lm\sp2+1)$. Thus both terms in the energy
are conserved independently. 
Their specific values are difficult to obtain in full 
generality; however, it is easy to calculate them for some specific 
solutions ({\it ie} specific choices of projectors $P$ and $R$ 
). 
Thus for $P=(p^\dg \otimes p)/|p|^2$ for $p=(1+|z|^2)(1,z)-2it(\br{z},-1)$
and $R=(q^\dg \otimes q)/|q|^2$ for $q=(1,z)$, where $z=x+iy$, we find 
\be
E_{2-\mbox{\small un}}=(8-4\lm b)\cdot 2\pi + {8\over \lm\sp2 +1}\cdot 0.
\label{resul}
\ee

Note that our energy (Hamiltonian) is different from the expression given in 
\cite{IZ}.
In fact both expressions have the meaning of energy and both are conserved.
The expression given in \cite{IZ} comes naturally when one considers
the energy momentum tensor, our expression (\ref{ham}) arises naturally
when one constructs the action for the model.
The existence of two, independent, Hamiltonians, is related to the
integrability of the model. In fact, integrable models often have many
independent Hamiltonians \cite{AC} and, not
surprisingly, this is the case here as well.

In conclusion - we have derived an action for the generalised chiral model.
This action contains a Wess-Zumino like term and it leads to a Hamiltonian 
which contains a similar contribution.
The equations of motion follow naturally from this action or from the
Hamiltonian using the standard Poisson brackets.

{\bf ACKNOWLEDGMENTS}
 
We thank R. S. Ward and B. Piette for helpful discussions.
TI acknowledges support from EU ERBFMBICT950035.

\end{document}